**FULL PAPER**    Open Access

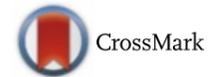

# Eastward-expanding auroral surges observed in the post-midnight sector during a multiple-onset substorm

Yoshimasa Tanaka[1,2*], Yasunobu Ogawa[1,2], Akira Kadokura[1,2], Noora Partamies[3,4], Daniel Whiter[5], Carl-Fredrik Enell[6], Urban Brändström[7], Tima Sergienko[7], Björn Gustavsson[8], Alexander Kozlovsky[9], Hiroshi Miyaoka[1,2] and Akimasa Yoshikawa[10,11]

**Abstract**

We present three eastward-expanding auroral surge (EEAS) events that were observed intermittently at intervals of about 15 min in the post-midnight sector (01:55–02:40 MLT) by all-sky imagers and magnetometers in northern Europe. It was deduced that each surge occurred just after each onset of a multiple-onset substorm, which was small-scale and did not clearly expand westward, because they were observed almost simultaneously with Pi 2 pulsations at the magnetic equator and magnetic bay variations at middle-to-high latitudes associated with the DP-1 current system. The EEASs showed similar properties to omega bands or torches reported in previous studies, such as recurrence intervals of about 15 min, concurrence with magnetic pulsations with amplitudes of several tens of nanotesla, horizontal scales of 300–400 km, and occurrence of a pulsating aurora in a diffuse aurora after the passage of the EEASs. Furthermore, the EEASs showed similar temporal evolution to the omega bands, during which eastward-propagating auroral streamers occurred simultaneously in the poleward region, followed by the formation of north-south-aligned auroras, which eventually connected with the EEASs. Thus, we speculate that EEASs may be related to the generation process of omega bands. On the other hand, the EEASs we observed had several properties that were different from those of omega bands, such as greater eastward propagation speed (3–4 km/s), shorter associated magnetic pulsation periods (4–6 min), and a different ionospheric equivalent current direction. The fast eastward propagation speed of the EEASs is consistent with the speed of eastward expansion fronts of the substorm current wedge reported in previous studies. The difference in the ionospheric current between the EEASs and omega bands may be caused by a large temporal variation of the surge structure, compared with the more stable wavy structure of omega bands.

## Background

It is well known that the form of an aurora depends on the local time and substorm phase (e.g., Akasofu 1964; Partamies et al. 2015). Large-scale wavy structures at the poleward boundary of a diffuse aurora, the so-called omega bands or torches (sometimes called tongues), are typical auroral phenomena observed in the post-midnight sector of the auroral zone, in particular during the recovery phase of a substorm (e.g., Akasofu and Kimball 1964; Oguti 1981; Opgenoorth et al. 1983; Safargaleev et al. 2005). They have a horizontal scale size of several hundred to more than 1000 km, which is often greater than the field of view (FOV) of a ground-based all-sky imager, and a lifespan of several tens of minutes. Since omega bands and torches propagate eastward at a speed of 0.4–2 km/s, corresponding to the E × B drift, they pass over the FOV of an all-sky imager in about 5–20 min. Thus, it is difficult to investigate their temporal evolution with only a single

* Correspondence: ytanaka@nipr.ac.jp
[1]National Institute of Polar Research, 10-3, Midori-cho, Tachikawa-shi, Tokyo 190-8518, Japan
[2]Department of Polar Science, School of Multidisciplinary Sciences, SOKENDAI (The Graduate University for Advanced Studies), 10-3 Midori-cho, Tachikawa, Tokyo 190-8518, Japan
Full list of author information is available at the end of the article imager, and the process from generation to extinction is still not completely understood.

Global auroral imager data from satellites have provided information on the evolution of omega bands and torches clarifying the relationship between poleward boundary intensifications (PBIs), auroral streamers, north-south (N-S)-aligned auroras, and omega bands. The intensification of the poleward boundary of the auroral oval and subsequent auroral streamers is followed by the equatorward ejection of N-S-aligned auroral forms, which finally evolve into auroral torch and omega-band structures in the post-midnight sector (Henderson et al. 1998, 2002; Henderson 2012). The PBIs, auroral streamers, and N-S auroras are interpreted to be the ionospheric manifestation of earthward-directed bursty bulk flows (BBFs) that are created by the reconnection in the magnetotail during the substorm expansion phase (Angelopoulos et al. 1992, 1994). However, since the spatial and temporal resolutions of the satellite observation data are limited, it is necessary to analyze data from ground-based imager and magnetometer networks to study the spatiotemporal development of these auroral phenomena in more detail.

In this paper, we present three meso-scale (200–500 km) auroral surge events observed in the post-midnight sector (around 01:55–02:40 magnetic local time (MLT); MLT ~ UT + 2 h at Tromsø (TRO)) by ground-based all-sky imagers and magnetometers. The auroral surges occurred intermittently at intervals of about 15 min just after the expansion onsets of a substorm and propagated eastward. We investigate the characteristics of the eastward-expanding auroral surges (hereafter referred to as EEASs) in detail and compare them with those of omega bands and other auroral phenomena in the post-midnight sector reported by previous studies.

## Methods

We conducted a campaign of auroral observations in Fennoscandia using multiple imagers and the European Incoherent Scatter (EISCAT) UHF radar from March 5 to March 9, 2013. The campaign was a collaboration by scientists from the Swedish Institute of Space Physics (IRF), the Finnish Meteorological Institute (FMI), the Sodankylä Geophysical Observatory (SGO), University of Tromsø, and the National Institute of Polar Research (NIPR). For this campaign, four CCD imagers of the Auroral Large Imaging System (ALIS) (Brändström 2003), two all-sky electron-multiplying CCD (EMCCD) imagers of the MIRACLE network (Syrjäsuo et al. 1998; Sangalli et al. 2011), and an all-sky EMCCD imager and all-sky Watec imagers (AWIs) installed at the EISCAT radar site in Tromsø (TRO), Norway, by the NIPR (Ogawa et al. 2013) were operated. The goal of this campaign was to reconstruct the three-dimensional (3D) distribution of the auroral emission and horizontal distribution of the energy spectra of precipitating electrons from the multiple auroral images using the tomographic inversion technique (Aso et al. 1998; Gustavsson 1998; Tanaka et al. 2011). In this paper, however, we focus on the characteristics of three auroral surges observed during the substorm on March 8–9, 2013. The results of the tomography analysis will be presented in another paper.

Figure 1 shows the location of the all-sky imagers and the IMAGE magnetometer chain (Lühr et al. 1998). Table 1 lists the geographical and the altitude-adjustment corrected geomagnetic (AACGM) coordinates of the observation stations used in this paper. The three EMCCD imagers were equipped with narrow bandpass filters (about 2 nm wide) to measure $N_2^+$ emission at 427.8 nm at 2-s intervals. The AWIs at TRO measured the emissions at 428, 558, and 630 nm using wider bandpass filters (about 100 nm) at 1-s intervals and white light at 0.5-s intervals. The sampling rate of the IMAGE magnetometers is 10 s; thus, we present all the optical data at 10-s intervals to correspond to the magnetometer data. Unfortunately, the EISCAT UHF radar and the ALIS imagers were not running when the auroral surges occurred.

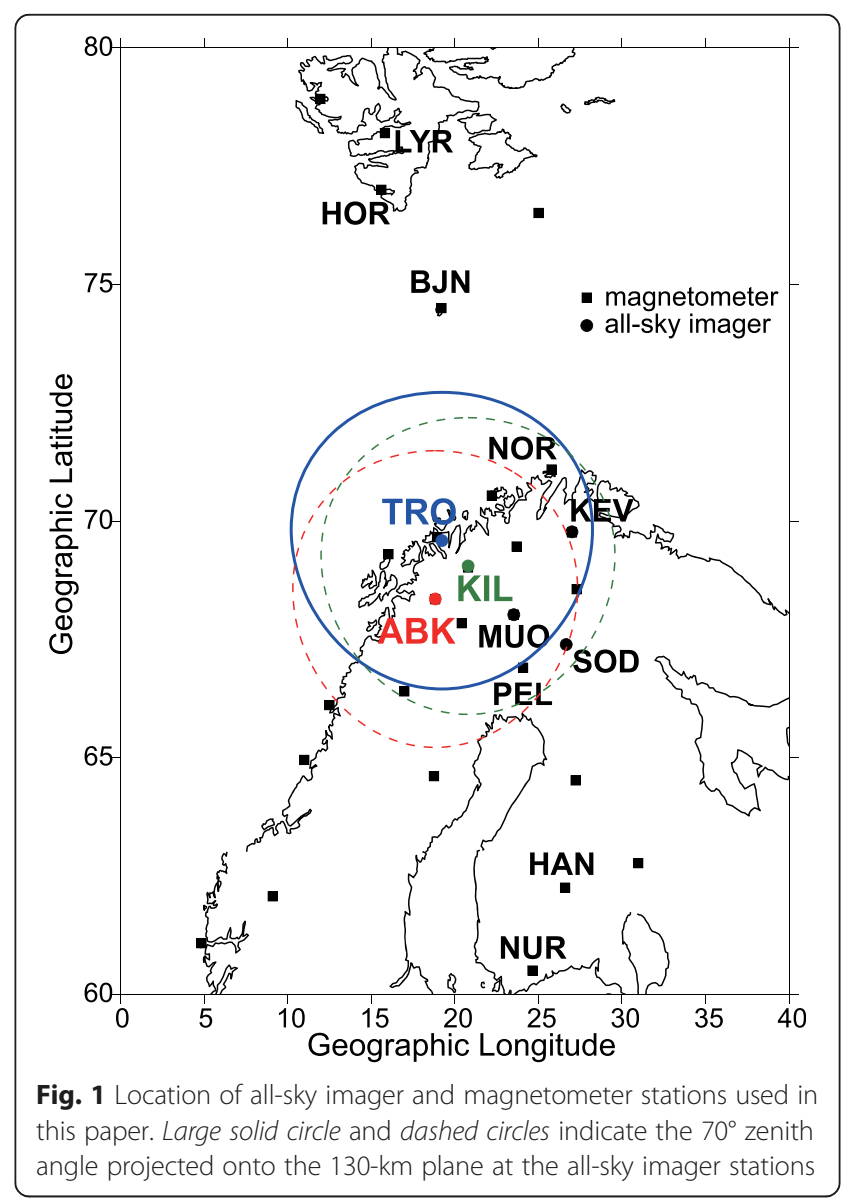

**Fig. 1** Location of all-sky imager and magnetometer stations used in this paper. *Large solid circle* and *dashed circles* indicate the 70° zenith angle projected onto the 130-km plane at the all-sky imager stations

**Table 1** Geographical and altitude-adjustment corrected geomagnetic (AACGM) coordinates of stations used in this paper

| Abbrev. | Station name | Geogr. lat. | Geogr. lon. | AACGM lat. | AACGM lon. |
|---|---|---|---|---|---|
| HOR | Hornsund | 77.00 | 15.60 | 73.94 | 110.48 |
| BJN | Bear Island | 74.50 | 19.20 | 71.27 | 108.94 |
| NOR | Nordkapp | 71.09 | 25.79 | 67.50 | 109.94 |
| SOR | Sørøya | 70.54 | 22.22 | 67.13 | 106.73 |
| KEV | Kevo | 69.76 | 27.01 | 66.10 | 109.73 |
| TRO (mag.) | Tromsø | 69.66 | 18.94 | 66.44 | 103.47 |
| TRO (imager) | Tromsø | 69.58 | 19.23 | 66.34 | 103.62 |
| MAS | Masi | 69.46 | 23.70 | 65.96 | 106.93 |
| AND | Andenes | 69.30 | 16.03 | 66.25 | 100.96 |
| KIL | Kilpisjärvi | 69.02 | 20.79 | 65.67 | 104.34 |
| IVA | Ivalo | 68.56 | 27.29 | 64.88 | 109.03 |
| ABK | Abisko | 68.35 | 18.82 | 65.11 | 102.30 |
| MUO | Muonio | 68.02 | 23.53 | 64.51 | 105.70 |
| KIR | Kiruna | 67.84 | 20.42 | 64.49 | 103.15 |
| PEL | Pello | 66.90 | 24.08 | 63.34 | 105.36 |
| JCK | Jäckvik | 66.40 | 16.98 | 63.21 | 99.46 |
| DON | Dønna | 66.11 | 12.50 | 63.21 | 95.79 |
| HAN | Hankasalmi | 62.25 | 26.60 | 58.47 | 104.85 |
| NUR | Nurmijärvi | 60.50 | 24.65 | 56.74 | 102.47 |
| AAB | Adis Ababa | 9.04 | 38.77 | 0.18 | 110.47 |

The AACGM coordinates were calculated for the year 2013 at 0-km altitude

## Results

Figure 2 shows a north-south keogram (i.e., geomagnetic north-south slices through the time series of AWI images) from TRO and the north-south ($X$) component of the magnetic field observed at eight stations of the IMAGE chain during the period from 22:00 to 02:00 UT (from 00:00 to 04:00 MLT at TRO) on March 8–9, 2013. Magnetometer data from Adis Ababa (AAB), which is located at the magnetic equator and belongs to the MAGDAS/CPMN managed by Kyushu University (Yumoto and MAGDAS Group 2006, 2007), has been filtered in the period range of Pi 2 pulsations (40–150 s) and is shown at the bottom of this figure. A gradual negative bay corresponding to an enhancement of the westward electrojet current can be seen at the stations located in the auroral zone (NOR, TRO, and KIL) from 22:50 to 01:10 UT (00:50–03:10 MLT). Variations resembling relatively sharp negative bays with small amplitudes (30–50 nT) can be seen at NOR at 00:00, 00:15, and 00:30 UT. A localized negative variation was also observed at Bear Island (BJN) at 00:37 UT. Simultaneously with the negative bays, positive bay variations with amplitudes less than 20 nT were observed at the mid-latitude stations (HAN and NUR) and the higher latitude stations (HOR and BJN). These negative and positive bays are consistent with a DP-1 (or substorm current wedge) current system, which typically occurs

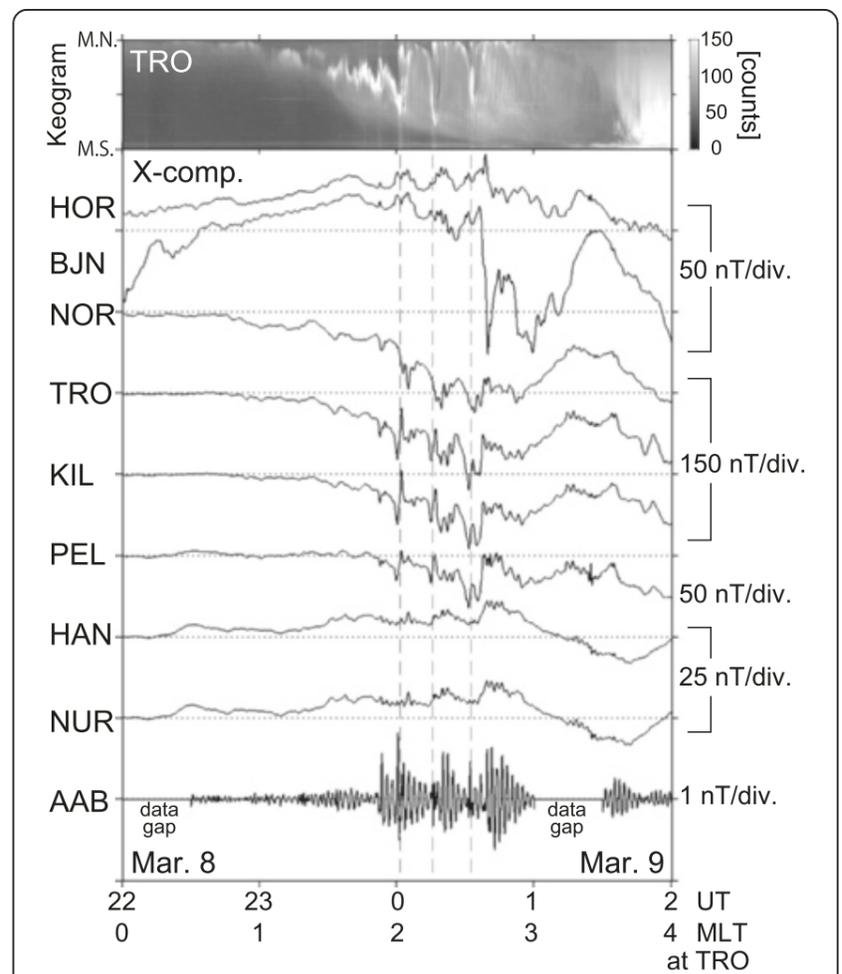

**Fig. 2** North-south keogram from TRO and north-south ($X$) component of the magnetic field from the magnetometer network. *Top panel* shows a north-south keogram obtained from the AWI imager data for the period from 22:00 to 02:00 UT on March 8–9, 2013. In the *bottom panel*, the *upper eight lines* show the magnetic variations detected by the IMAGE magnetometer chain. The *bottom line* shows filtered data in the period range of Pi 2 pulsations (40–150 s) at AAB, which belongs to the MAGDAS/CPMN and is located at the magnetic equator. *Three vertical dashed lines* indicate the time when the front of the eastward-expanding auroral surges reached the zenith at TRO

during the substorm expansion phase (see Kepko et al. 2014 and references therein). The magnetic variations associated with the DP-1 current system were also confirmed in the $Y$ component at mid-latitude (not shown here). Palin et al. (2015) showed that even small and localized DP-1 current systems seem to be a consequence of plasma sheet activation in the near-Earth tail, which is associated with BBFs or depolarization fronts. Three or four successive Pi 2 pulsations were observed at AAB in the interval 23:50–01:00 UT (01:50–03:00 MLT); they are usually used as indicators of a substorm onset. Three vertical dashed lines indicate the times when the fronts of the eastward-expanding auroral surges (which will be shown below) reached the zenith at TRO. In the keogram, each EEAS is characterized as an equatorward movement of a discrete arc and subsequent rapid poleward expansion of the diffuse auroral region. These times coincide with the occurrence of Pi 2 pulsations. Small but similar optical and magnetic variations can also be identified at 23:50 UT.

Both solar wind and geomagnetic conditions were moderate during this interval. According to ACE satellite data,

the solar wind speed and proton density were about 340 km/s and 8–10/cm$^3$, respectively. The Bz and By components of the interplanetary magnetic field (IMF) were between −4 and −1 nT and between −2 and 0 nT, respectively. The K$_P$ index was between 1− and 2+ in this interval, and the provisional AE index was about 100 nT at 00:40 UT on March 8.

Figure 3 exhibits a sequence of all-sky images for the three EEAS events observed at TRO. These color images were composed from the 428-, 558-, and 630-nm images taken by the AWIs. We first focus on the second event (Fig. 3b). Before the auroral surge comes into the FOV of the imager, diffuse aurora including pulsating aurora covers the FOV (00:14:30 UT). Then, the poleward edge of the diffuse aurora gradually moved equatorward (00:14:40–00:15:00 UT). An auroral streamer (marked by A.S. in the figure) emerged at the western edge of the FOV, at a higher latitude than the diffuse aurora (00:15:00 UT), and propagated eastward. Almost simultaneously with the auroral streamer, the front of an auroral surge (i.e., EEAS) propagated eastward along the poleward edge of the diffuse aurora and spread over most of the FOV (00:15:00–00:16:20 UT). In its poleward region, the auroral streamer formed spirals and branched at several places (e.g., marked by arrows at 00:15:50 and 00:16:10 UT). North-south-aligned auroras emerged behind the vortex structure (NS in the figure at 00:16:20 and 00:16:40 UT) and finally connected the branched streamer with the EEAS. From these all-sky images, we could not determine whether N-S-aligned auroras emerged from any of the auroral streamers or the auroral surge. The vortex structure of the surge finally converged on the eastern side of the FOV (00:16:50 UT). After the passage of the EEAS, the diffuse aurora covered the FOV again, and a pulsating aurora was observed in the diffuse aurora (after 00:17:30 UT). The dynamic behavior of the EEAS is very similar for all three events, which occurred intermittently at intervals of about 15 min. Thus, the EEAS can be regarded as a repeatedly occurring phenomenon during the substorm.

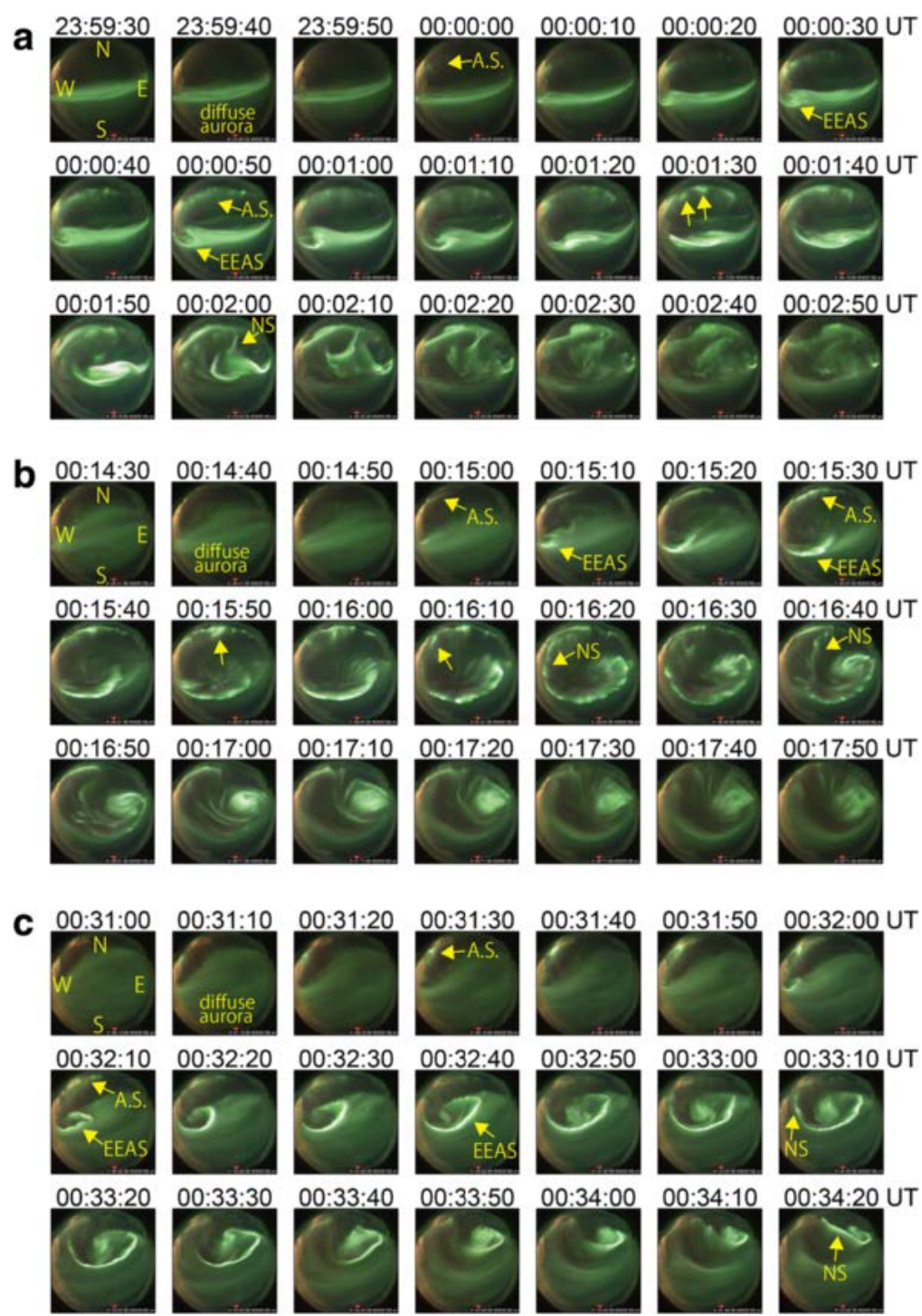

**Fig. 3** Sequence of all-sky images for three eastward-expanding auroral surges observed at TRO. **a** The first EEAS event detected in the interval 23:59:30–00:02:50 UT. **b** The second event in the interval 00:14:30–00:17:50 UT. **c** The third event in the interval 00:31:00–00:34:20 UT. These color images were composed from the 428-, 558-, and 630-nm images taken by the all-sky Watec imagers (AWIs). Magnetic north and east correspond to the *top* and *right*, respectively

We also examined optical and magnetic data taken at other stations, such as Narsarsuaq (NRSQ; 61.16°N, 314.56°E) in Greenland, Husafell (HUS; 64.67°N, 338.97°E) and Tjornes (TJO; 66.20°N, 342.88°E) in Iceland, Kevo (KEV) and Sodankylä (SOD) in Finland, and Longyearbyen (LYR) in Svalbard, Norway. It was overcast at HUS and TJO, although an auroral brightening occurred at these stations at nearly the same time as the EEASs. Since the magnetic variations observed at TJO and HUS were similar to those at TRO, the substorm expanded at least to this meridian. We identified an auroral arc at NRSQ located 4 h west of TRO, which may correspond to the arc at the poleward boundary of the diffuse aurora at TRO. The temporal variation of the luminosity of the arc was synchronized with the EEASs, but the arc at NRSQ was more stable compared with the dynamic behavior of the EEASs. In addition, no clear negative bay was observed in the $X$ component of the magnetic field at NRSQ. Thus, we concluded that this multiple-onset substorm was small-scale and did not expand to the premidnight sector. At KEV and SOD, the behavior of the EEASs is very similar to that at TRO because of the close location. On the other hand, the imager at LYR barely detected the EEASs at the southern edge of the FOV and did not provide additional information on the auroral activity in the poleward region.

Figure 4 shows an expanded view of the magnetic variations at TRO and Abisko (ABK) from 23:50 to 00:45 UT (from 01:50 to 02:45 MLT) on March 8–9. Systematic temporal variations can be clearly seen in all components of the magnetic field at times when the EEASs were observed. These variations are marked by dashed-dotted lines. The magnetic variation at TRO shows negative-positive excursions in the $X$ component, positive-negative-positive excursions in the east-west ($Y$) component, and positive-negative excursions in the vertical ($Z$) component for all the EEASs. The periods of the magnetic variations shown by the dashed-dotted lines are about 4–6 min. The peak-to-peak amplitudes of the magnetic variations are between 10 and 100 nT. The magnetic variation at ABK is similar to that at TRO, but it is anti-phase with respect to that at TRO only in the $Z$ component, which implies that the east-west component of

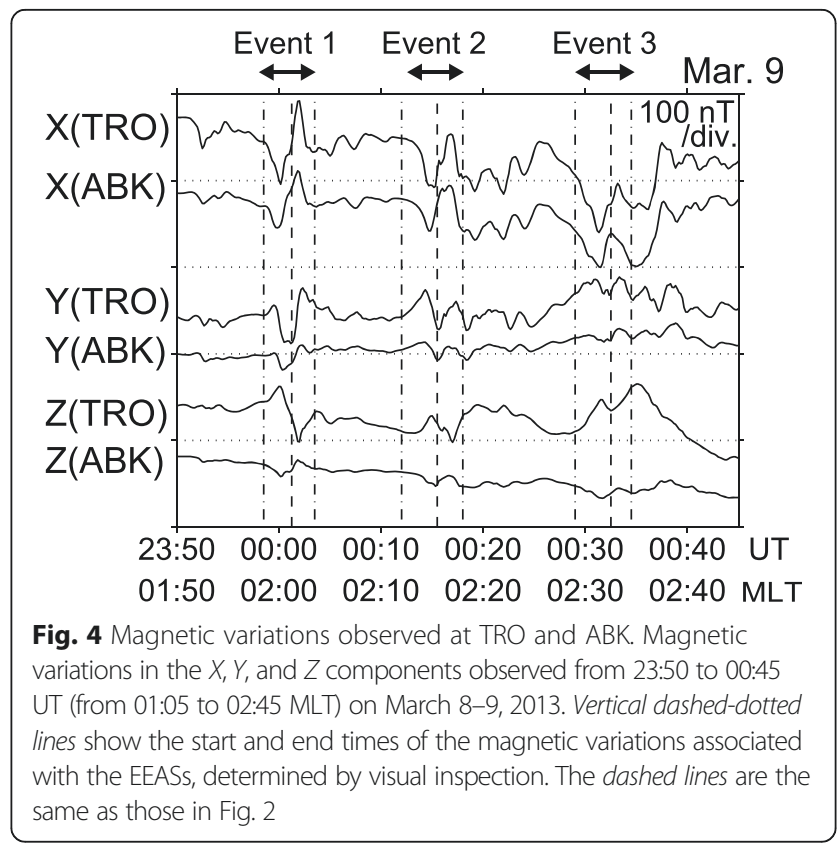

**Fig. 4** Magnetic variations observed at TRO and ABK. Magnetic variations in the $X$, $Y$, and $Z$ components observed from 23:50 to 00:45 UT (from 01:05 to 02:45 MLT) on March 8–9, 2013. *Vertical dashed-dotted lines* show the start and end times of the magnetic variations associated with the EEASs, determined by visual inspection. The *dashed lines* are the same as those in Fig. 2

the ionospheric current associated with the EEASs has maximal intensity between TRO and ABK.

Figure 5 shows the ionospheric equivalent current associated with the EEASs, plotted with the auroral images from TRO mapped onto a horizontal plane at a 105-km altitude. The ionospheric currents were derived from the IMAGE magnetometer data by a simple 90° clockwise (CW) rotation. The magnetic perturbations associated with the EEASs have been extracted by subtracting a linear trend from the raw data within each interval of the EEAS, as shown in Fig. 4. It should be noted that if we do not remove the trend, the ionospheric current is always roughly westward because the $X$ component is always negative during these intervals (Fig. 2). Again, we explain the temporal variation of the ionospheric current using the second event (Fig. 5b). Before the arrival of the EEAS, the ionospheric current was roughly westward along the auroral arc and showed a CW rotation centered at a higher latitude than the visible aurora (00:14:50 UT). The current vectors rapidly rotated clockwise as the EEAS passed over the FOV of the imager. The current system formed a counterclockwise (CCW) rotation along the auroral vortex structure or around the front of the surge when the center of the EEAS reached the zenith at TRO (00:16:20 UT). At 00:16:20 UT, the longitudinal size of the EEAS can be estimated to be about 400 km or 10° in longitude. When the size of the vortex became smaller at 00:17:10 UT, the CCW rotation of the ionospheric current system is evident around the vortex. This temporal variation in the ionospheric current is very similar for the other two EEAS events. It is well established that if the ionospheric conductivity is horizontally uniform, the rotational (divergence-free) ionospheric current that is observable on the ground is the Hall

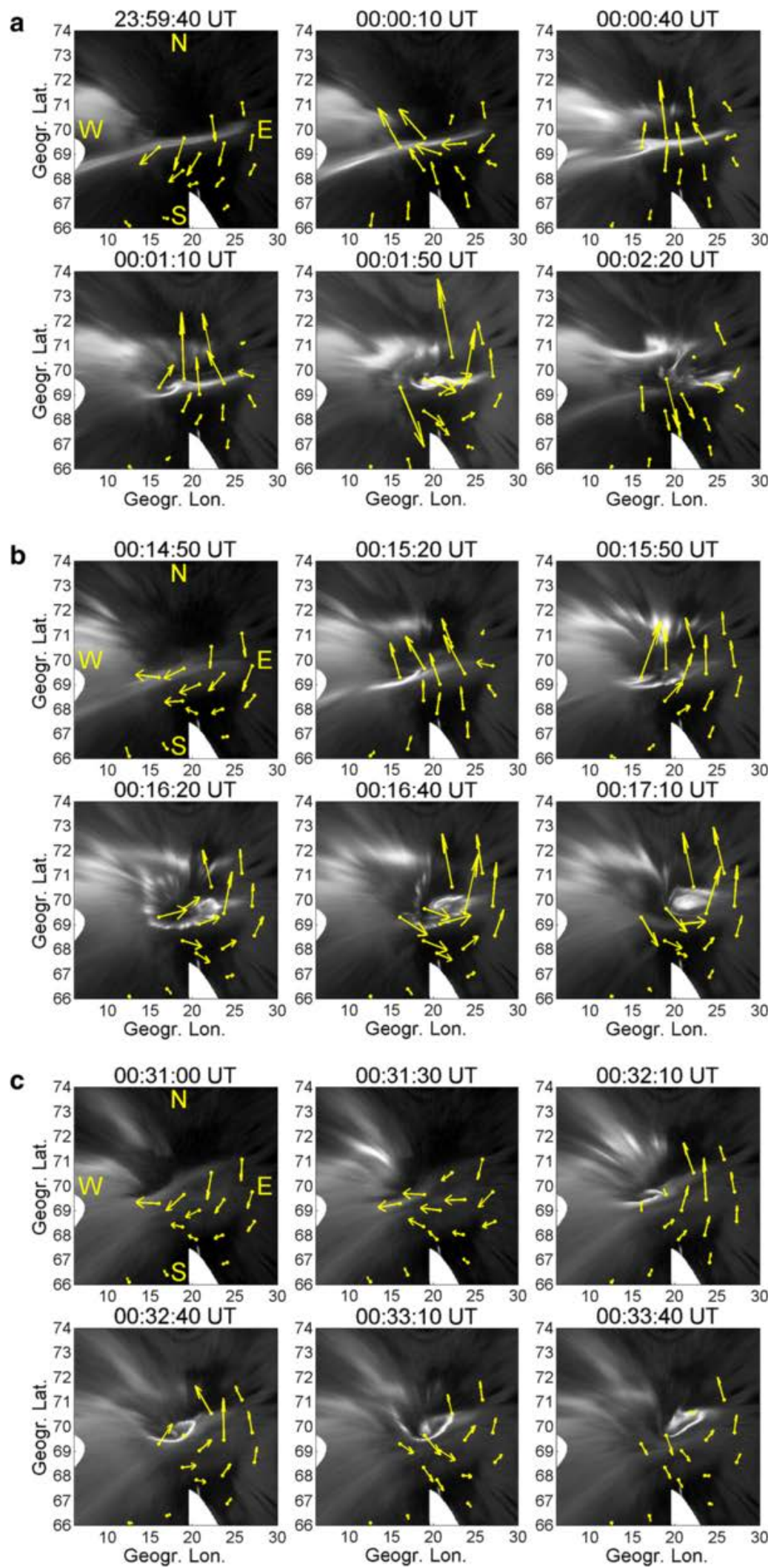

**Fig. 5** Ionospheric equivalent current system associated with EEASs. **a** Temporal evolution of the ionospheric equivalent currents for the first EEAS event, **b** second event, and **c** third event. The ionospheric equivalent current was derived from the magnetic perturbations associated with the EEASs by a 90° clockwise rotation and is plotted with the auroral images mapped onto a horizontal plane at a 105-km altitude

current, which flows CW (CCW) along the equipotential lines around the positive (negative) potential at high latitude in the Northern Hemisphere. Although the ionospheric current pattern is deformed because of non-uniform ionospheric conductivity, it is possible to deduce that the CW and CCW rotations in the ionospheric equivalent current correspond to a positive and negative ionospheric electrical potential, respectively, which are primarily caused by downward and upward field-aligned currents (FACs) (c.f., Glassmeier 1984).

The speed of the eastward expansion of the surges was estimated from the temporal variation of the auroral images mapped onto a horizontal plane at the 105-km altitude. The propagation speed was about 3, 4, and 3 km/s across the zenith of TRO for the first, second, and third event, respectively. The assumed altitude of 105 km is reasonable, because we applied tomography analysis to the 428-nm images taken by the three EMCCD imagers and confirmed that the maximal peaks of the emission were located around 95- to 115-km altitude (not shown here).

Furthermore, we calculated the footprints of the EEASs on the equatorial plane of the magnetosphere using the Tsyganenko 96 model. We found that the EEASs are mapped to a radial distance of 5–7 $R_E$, which is close to the footprints of torches (Tagirov 1993), and the propagation speed at the equatorial magnetosphere is 40–60 km/s.

## Discussion
### Comparison with omega bands

The large-scale vortex-like structure of the EEASs at the poleward edge of the diffuse aurora appears to be similar to the undulations of omega bands. In addition to the spatial structure, both have some similar properties, such as (1) occurrence in the post-midnight sector, (2) eastward propagation, (3) concurrence with magnetic pulsations with amplitudes of 10 to several 100 nT, (4) periodic recurrence with intervals of about 5–40 min, (5) horizontal scale sizes of several hundred kilometers, and (6) existence of a pulsating aurora in the diffuse aurora (Opgenoorth et al. 1983; Amm et al. 2005; Vanhamäki et al. 2009, and references therein). As for item (3), omega bands are in general accompanied by Ps 6 magnetic pulsations with an amplitude of 10 to more than 500 nT (Saito 1978), which are more dominant in the *Y* and *Z* components. However, we note that our EEASs show strong pulsations in the *X* component, which would be somewhat unusual for omega bands. Item (4) is generally accepted for omega bands because Ps 6 pulsations have periods of 5–40 min. The occurrence interval of about 15 min for the EEASs is equal to the typical average period of omega bands. The typical longitudinal as well as latitudinal extent of omega bands is about 400–500 km, which is comparable to that of the EEASs (item (5)). Regarding item (6), the pulsating aurora is commonly observed at the center of the torches, i.e., the diffuse auroral region equatorward of the omega bands (e.g., Oguti 1981).

On the other hand, there are some differences in the properties between the EEASs and the omega bands as follows: (a) It is deduced from the simultaneous observations of Pi 2 pulsations and the magnetic bay variations associated with the DP-1 current system that each EEAS was observed just after each onset of a multiple-onset substorm, whereas omega bands are usually observed during the substorm recovery phase. (b) The propagation speed of the EEASs (about 3–4 km/s) is faster than that of omega bands (0.4–2 km/s). (c) The period of the magnetic pulsations accompanying the EEASs (4–6 min) is shorter than that of the omega bands (5–40 min). (d) The direction of the ionospheric equivalent current of the EEASs appears to be different from that of omega bands.

As for item (d), omega bands have a systematic and relatively stable ionospheric equivalent current system, which is caused by the large-scale wavy structure along the east-west direction. The ionospheric current shows a CCW rotation around the torch, corresponding to the upward FAC and negative potential, and a CW rotation around the dark region (i.e., inside the inverted capital Greek letter omega ($\Omega$)), corresponding to the downward FAC and positive potential (Opgenoorth et al. 1983; Buchert et al. 1988; Wild et al. 2000; Amm et al. 2005; Vanhamäki et al. 2009). On the other hand, the ionospheric equivalent current of the EEASs indicates a quite dynamic behavior. It showed a CCW rotation around the auroral vortex structure at 00:16:20 UT for the second event (Fig. 5b), which corresponds to the structure of the inverted letter $\Omega$ of the omega bands; namely, the direction of the rotation is opposite to that of the omega bands. It is expected from the CCW rotation that the upward FAC and negative potential existed inside the auroral vortex, although they depend on the distribution of the ionospheric conductivity. However, for the case of the third event at 00:33:10 UT (Fig. 5c), the CCW rotation in the ionospheric current can be seen at the front of the surge. The difference in the ionospheric equivalent current of the EEAS from the omega bands (and from case to case) may be attributed to a large temporal variation in the surge structure, in contrast to the stable, well-defined wavy structure of omega bands; such a large temporal variation in the surge yields a variation in spatial distribution of the FACs and non-uniform ionospheric conductivity, which causes a deformation of the ionospheric current pattern.

### Interpretation of the EEASs

Since the EEASs coincided quite precisely with Pi 2 pulsations near the magnetic equator and magnetic bay variations associated with the DP-1 current system, it is reasonable to assume that they occurred just after substorm onsets. Nakamura et al. (1993) investigated the auroral evolution during the substorm expansion phase using all-sky TV data and demonstrated that the discrete auroral structures within the poleward expanding bulge developed not only westward but also eastward and equatorward. The eastward-propagating aurora and the north-south-aligned aurora eventually develop into a diffuse and pulsating aurora after the expansion. They suggested that these discrete auroral structures in the bulge should be attributed to the plasma streamlines in the magnetosphere. The auroral streamers in the poleward region and the N-S auroras shown in the present study (Fig. 3) may be associated with the eastward-propagating aurora and the N-S aurora shown by Nakamura et al. (1993), respectively.

Global auroral imager data from satellites have provided evidence that the poleward boundary intensifications (PBIs) and subsequent auroral streamers are followed by equatorward emergence of N-S-aligned auroras, which finally evolve into torches and omega bands (Henderson 2012). Although the PBIs/auroral streamers (or N-S auroras) are not necessarily associated with substorm onset, they can be launched from the poleward edge of the poleward expanding auroral substorm bulge that is fully embedded within the closed field line region; such PBIs/auroral streamers associated with the substorm expansion may cause the auroral streamers accompanying the EEASs. We note that EEASs have not been reported from satellite observations yet, which may be because the spatial and temporal resolutions of the global imagers on board the satellites were too low to detect them. In addition, there is a possibility that N-S auroras emerged from EEASs located in the equatorward region.

Phenomena similar to the EEASs have been reported by Yeoman and Lühr (1997) and Yeoman et al. (1998) using the CUTLASS Finland HF radar and the magnetometer chain, namely the ionospheric equivalent current vortices observed during the substorm expansion phase. They found that at least five current vortices, showing alternating senses of rotation (i.e., CW and CCW), propagated eastward during the expansion phase. The vortices are characterized by a duration of ~5 min, a repetition period of ~8 min, an eastward propagation speed of <6 km, and an azimuthal extent of about 1000 km. These characteristics of the current vortices are consistent with those of EEASs. To our knowledge, this is the only paper that has reported phenomena corresponding to the EEASs.

There have been several studies on the azimuthal expansion speed of the initial brightening arcs (Shiokawa et al. 2009; Ogasawara et al. 2011), although the estimated footprints of the arcs in the magnetosphere are located more tailward than those of the EEASs. Ogasawara et al. (2011) investigated the azimuthal auroral expansion and its magnetospheric counterpart using data from THEMIS all-sky imagers and multiple spacecraft and found that the average speed of the leading edge of the eastward auroral expansion was 5.3 km/s, which is comparable to the expansion speed of the EEASs. They also demonstrated that the speed mapped onto the equatorial plane of the magnetosphere (162 km/s) was comparable to the averaged azimuthal plasma (E × B) flow speed observed at a radial distance of ~12 $R_E$ by the spacecraft (112–139 km/s).

The azimuthal propagation speed of the substorm expansion at geosynchronous orbit ($L = 6.6$ $R_E$) was derived from the time difference between the magnetic field dipolarization onset observed by the GOES 8 and 9 satellites and the auroral breakup observed by the Polar ultraviolet imager (Liou et al. 2002). The typical eastward propagation speed of the dipolarization was ~60 km/s, which is consistent with the result from the present study. Thus, it is possible that the eastward expanding auroral surges and eastward expansion fronts of a substorm current wedge are related.

The fast propagation of the vortices may be alternatively explained by a surface wave on a plasma boundary in the closed magnetosphere (Maltsev and Lyatsky 1984). This model was originally suggested for Pi 2 pulsations and east-west motions of auroral riometer absorption bays and was later used to describe fast (4–6 km/s) azimuthal propagation of dayside traveling convection vortices (Lyatsky et al., 1999). In the frame of this model, the fast motion of auroral structures is a manifestation of the propagation of the magnetospheric wave, rather than the flow of magnetospheric plasma.

Saito (1978) showed an example of Ps 6 pulsations (considered to coincide with omega bands) that started simultaneously with a poleward expansion just after an auroral breakup. They also indicated that the Ps 6 onset coincided with the Pi 2 onset in the midnight sector and showed a linear time lag after the Pi 2 onset in the dawn and dusk sectors. Furthermore, Connors et al. (2003) provided several examples in which Ps 6 pulsations occurred at or very near the onset of a substorm expansion phase, a pseudo-breakup, or a poleward boundary intensification. Wild et al. (2011) also presented an example of omega bands observed just after substorm onset in the midnight sector, dawnward of the onset region. Considering these previous studies and the analogy between the EEASs and omega bands, we speculate that the EEASs may be related to the generation process of the omega bands, and quite a few omega bands observed in the recovery phase may originate in the eastward expansion of the auroral bulge. The faster eastward propagation of the EEASs (3–4 km/s) compared to the omega bands, resulting in a shorter period of the magnetic pulsations than that of Ps 6 pulsations, may be explained by the movement of eastward expansion fronts of the substorm current wedge. The difference in the ionospheric equivalent current between the EEASs and omega bands may be attributed to a large temporal variation in the surge structure, in which the distributions of the FACs and the ionospheric conductivity vary from moment to moment, in contrast to the more stable, well-defined wavy structure of omega bands. Unfortunately, we cannot confirm the abovementioned hypothesis because of a lack of data from global multi-imager networks as well as satellites. It is necessary to statistically analyze data from the large-scale multi-imager and multi-magnetometer networks such as the THEMIS GBO in order to clarify the spatiotemporal variation of the EEASs; this will be explored in future studies.

## Conclusions

Three eastward-expanding auroral surges (EEASs) were observed intermittently at intervals of about 15 min in the post-midnight sector (1:55–2:40 MLT) by all-sky imagers and magnetometers in northern Europe. From the simultaneous observation of Pi 2 pulsations at the magnetic equator and magnetic bay variations at middle-to-high latitudes associated with the DP-1 current system, it is deduced that each EEAS occurred just after each onset of a multiple-onset substorm, which was small-scale and did not expand to the premidnight sector. The EEASs have similar characteristics to omega bands or torches, such as recurrence intervals of about 15 min, concurrence with magnetic pulsations with an amplitude of several tens of nanotesla, horizontal scales of 300–400 km, and occurrence of pulsating auroras in the diffuse auroral region after the passage of the EEASs. In addition to these similarities, the temporal evolution of the EEASs, showing that the eastward-propagating auroral streamers simultaneously occur in the poleward region and eventually connect with the EEASs, is similar to that of omega bands. Thus, we speculate that the EEASs may be related to the generation process of the omega bands. On the other hand, there are some differences in the characteristics between the EEASs and omega bands, i.e., faster eastward propagation (3–4 km/s), resulting in a shorter period of magnetic pulsations (4–6 min), and a different direction for the ionospheric equivalent current. The fast eastward propagation of the EEAS may be interpreted as the movement of eastward expansion fronts of the substorm current wedge. The difference in the ionospheric current between the EEASs and omega bands may be caused by a large temporal variation in the surge structure, compared with the more stable wavy structure of omega bands.

### Abbreviations
AACGM: altitude-adjusted corrected geomagnetic; ALIS: Auroral Large Imaging System; AWI: all-sky Watec imager; CCW: counterclockwise; CW: clockwise; EEAS: eastward-expanding auroral surge; EISCAT: European Incoherent Scatter Scientific Association; EMCCD: electron-multiplying CCD; FAC: field-aligned current; PBI: poleward boundary intensification.

### Competing interests
The authors declare that they have no competing interests.

### Authors' contributions
YT conducted the auroral campaign observation, analyzed the observed data, and prepared the manuscript. YO operated the EISCAT UHF radar and all-sky imagers at TRO. AK (NIPR) operated the all-sky imagers at Skibotn. NP, CFE, and DW operated the EMCCD all-sky imagers of the MIRACLE network. UB and TS operated the ALIS imagers. NP calculated the ionospheric equivalent current. AY provided the magnetometer data at AAB. YO, AK (NIPR), NP, BG, TS, AK (SGO), and HM joined and contributed to the discussion. All the authors read and approved the final manuscript.

### Acknowledgements
The MIRACLE network is operated as an international collaboration under the leadership of the Finnish Meteorological Institute. The IMAGE magnetometer data are collected as a joint European collaboration. The $K_P$ and provisional AE indices are provided by the World Data Center for Geomagnetism, Kyoto University. We thank the NASA National Space Science Data Center, the Space Physics Data Facility, and the ACE Principal Investigator, Edward C. Stone of the California Institute of Technology, for the use of ACE data. This work was supported by the NIPR Project KP-9 and also supported in part by the Inter-university Upper atmosphere Global Observation NETwork (IUGEONET) project (http://www.iugonet.org/en/) funded by the Ministry of Education, Culture, Sports, Science and Technology (MEXT), Japan. This work was also partly supported by the Transdisciplinary Research Integration Center, Research Organization of Information and Systems, Japan. The production of this paper was supported by an NIPR publication subsidy.

### Author details
[1]National Institute of Polar Research, 10-3, Midori-cho, Tachikawa-shi, Tokyo 190-8518, Japan. [2]Department of Polar Science, School of Multidisciplinary Sciences, SOKENDAI (The Graduate University for Advanced Studies), 10-3 Midori-cho, Tachikawa, Tokyo 190-8518, Japan. [3]The University Centre in Svalbard, 156, 9171 Longyearbyen, Norway. [4]Arctic Research, Finnish Meteorological Institute, Box 503, FI-00101 Helsinki, Finland. [5]School of Physics and Astronomy, University of Southampton, Southampton SO17 1BJ, UK. [6]EISCAT Scientific Association, Box 812, SE-981 28 Kiruna, Sweden. [7]Swedish Institute of Space Physics, Box 812, SE-981 28 Kiruna, Sweden. [8]Department of Physics and Technology, University of Tromsø, Tromsø, Norway. [9]Sodankylä Geophysical Observatory, Tähteläntie 62, FIN-99600 Sodankylä, Finland. [10]Department of Earth and Planetary Sciences, Faculty of Sciences, 33 Kyushu University, 6-10-1 Hakozaki, Higashi-ku, Fukuoka 812-8581, Japan. [11]International Center for Space Weather Science and Education, 53 Kyushu University, 6-10-1 Hakozaki, Higashi-ku, Fukuoka 812-8581, Japan.